\documentclass[12pt,preprint]{emulateapj}

\usepackage{natbib}
\usepackage{rotate}
\def\gtorder{\mathrel{\raise.3ex\hbox{$>$}\mkern-14mu
             \lower0.6ex\hbox{$\sim$}}}
\def\ltorder{\mathrel{\raise.3ex\hbox{$<$}\mkern-14mu
             \lower0.6ex\hbox{$\sim$}}}

\newcommand{\feh}{\hbox{$ [{\rm Fe}/{\rm H}]$}}

\shorttitle{Bulge RR Lyr population from the OGLE-III data}
\shortauthors{Pietrukowicz et al.}

\begin{document}

\title{THE OPTICAL GRAVITATIONAL LENSING EXPERIMENT:\\
ANALYSIS OF THE BULGE RR LYRAE POPULATION FROM THE OGLE-III DATA}

\author{P.~Pietrukowicz\altaffilmark{1},
A.~Udalski\altaffilmark{1},
I.~Soszy{\'n}ski\altaffilmark{1},
D.~M.~Nataf\altaffilmark{2},
{\L}.~Wyrzykowski\altaffilmark{1,3},
R.~Poleski\altaffilmark{1},
S.~Koz{\l}owski\altaffilmark{1},
M.~K.~Szyma{\'n}ski\altaffilmark{1},
M.~Kubiak\altaffilmark{1},
G.~Pietrzy{\'n}ski\altaffilmark{1,4},
and K.~Ulaczyk\altaffilmark{1},
}

\altaffiltext{1}{Warsaw University Observatory,
Al. Ujazdowskie 4, 00-478 Warszawa, Poland}
\altaffiltext{2}{Department of Astronomy, Ohio State University,
140 West 18th Avenue, Columbus, OH 43210, USA}
\altaffiltext{3}{Institute of Astronomy, University of Cambridge,
Madingley Road, Cambridge CB3 0HA, UK}
\altaffiltext{4}{Departamento de F\'isica, Universidad de Concepci{\'o}n,
Casilla 160-C, Concepci{\'o}n, Chile}

\begin{abstract}
We have analyzed the data on 16,836 RR Lyrae (RR Lyr) variables observed
toward the Galactic bulge during the third phase of the Optical Gravitational
Lensing Experiment (OGLE-III), which took place in 2001--2009. Using these
standard candles, we show that the ratio of total to selective extinction toward
the bulge is given by $R_I=A_I/E(V-I)=1.080\pm0.007$ and is independent of
color. We demonstrate that the bulge RR Lyr stars form a metal-uniform
population, slightly elongated in its inner part. The photometrically
derived metallicity distribution is sharply peaked at $\feh=-1.02\pm0.18$, with
a dispersion of 0.25~dex. In the inner regions ($|l|<3\degr$, $|b|<4\degr$)
the RR Lyr tend to follow the barred distribution of the bulge red
clump giants. The distance to the Milky Way center inferred from
the bulge RR Lyr is $R_0=8.54\pm0.42$~kpc. We report a break
in the mean density distribution at a distance of $\sim0.5$~kpc from
the center indicating its likely flattening. Using the OGLE-III data,
we assess that $(4$--$7) \times 10^4$ type ab RR Lyr variables should be
detected toward the bulge area of the on going near-IR VISTA Variables
in the Via Lactea (VVV) survey, where the uncertainty partially results
from the unknown RR Lyr spatial density distribution within $0.2$~kpc
from the Galactic center.
\end{abstract}



\section{INTRODUCTION}
\label{sec:introduction}

RR Lyrae (RR Lyr) variable stars are known as standard candles and useful
tracers of old populations. They play an essential role in our understanding of
the formation and evolution of the Galaxy. When multi-epoch light curves
are available, RR Lyr variables can be easily identified. A large number
of observations allows determination of their properties with high
precision. The derived parameters can be used to determine interstellar
extinction, metallicity, distance, and spatial distribution of the stars.
RR Lyr stars observed toward the bulge provide an independent determination
of the distance to the center of the Milky Way.

RR Lyr stars observed close to the central regions of the Milky Way
concentrate toward the Galactic center was first noticed
by \cite{1932BAN.....6..163V,1933BAN.....7...21V}.
Later, \cite{1946PASP...58..249B} found a strong predominance
of RR Lyr variables toward a relatively unobscured area today called
the Baade's window, indicating the presence of Population II stars in the
Milky Way center. By the early 1990s approximately 1000 RR Lyr
variables inhomogeneously distributed toward the bulge were known.

The number of new bulge RR Lyr variables has increased following the advent
of massive photometric surveys. \cite{1994AcA....44..317U, 1995AcA....45....1U,
1995AcA....45..433U, 1996AcA....46...51U, 1997AcA....47....1U}
published a catalog of over 3000 periodic variable stars detected in the
fields covered by the first phase of the Optical Gravitational Lensing
Experiment (OGLE-I). Two-hundred and fifteen of these objects were
classified as RR Lyr. Analysis of the OGLE-II data brought a much
larger list of 2713 RR Lyr (\citealt{2003AcA....53..307M}).
Later, using the same source of data, \cite{2006ApJ...651..197C}
prepared a catalog of 1888 fundamental mode RR Lyr stars (type RRab).

The MACHO microlensing project also observed a numerous sample of
RR Lyr stars toward the Galactic center. \cite{1998IAUS..184..123M},
using a sample of 1150 RRab and 550 RRc stars, showed that the spatial
distribution of RR Lyr between 0.3 and 3~kpc follows a power law with
an inclination of $-3.0$. \cite{1998ApJ...492..190A} examined the mean
colors and magnitudes of $\sim$1800 RR Lyr and found that the bulk of the
population is not barred. Only RR Lyr located toward the inner fields closer
to the Galactic center ($l<4\degr$, $b>-4\degr$) seem to follow the barred
distribution observed for red clump giants (RCGs) (\citealt{1994ApJ...429L..73}).
Recently, \cite{2008AJ....135..631K} analyzed photometric data on 3525 MACHO
RRab stars to assess the reddening toward the Galactic bulge. They derived
the selective extinction coefficient $R_{V,VR}=A_V/E(V-R)=4.3\pm0.2$,
which corresponds to the average value observed in the solar neighborhood
$R_{V,BV}=A_V/E(B-V)=3.1\pm0.3$  (\citealt{1989ApJ...345..245C,1999PASP..111...63F}).

The first metallicity measurements of bulge RR Lyr stars were made by
\cite{1976ApJ...210..120B} using the $\Delta S$ method
(\citealt{1959ApJ...130..507P}). Using 9 stars in the ~100 variable sample
of \cite{1963Sci...140..658I} list, they obtained $\langle\feh\rangle$ $=-0.65\pm0.15$~dex
and concluded that the stars are mildly metal-poor. \cite{1986A&A...169..111G} used
a sample of 17 bulge RR Lyr variables and found a wide range of iron abundances,
between $-1.8$ and $+0.1$~dex. Later, from spectra of 59 RRab and RRc variables,
\cite{1991ApJ...378..119W} determined an average metallicity of
$\langle\feh\rangle$ $=-1.0$~dex on the \cite{1984ApJS...55...45Z} metallicity scale.
In contrast to previous studies they found the metallicity distribution
to be sharply peaked, with a dispersion of only 0.16~dex. The most
recent metallicity estimate was performed by \cite{2008AJ....136.2441K}.
Based on photometric data for 2,435 MACHO RRab stars they derived
$\langle\feh\rangle$ $=-1.25$ (on the \citealt{1984ApJS...55...45Z} scale),
with a broad metallicity range from $\feh=-2.26$ to $-0.15$~dex.

The bulge RR Lyr stars have also often been used to measure the distance to the
Galactic center, under the assumption that the center of their population
corresponds to the center of the Milky Way. The first such measurement by
\cite{1951POMic..10....7B} yielded $R_0=8.7$~kpc. For more than half a century, various
investigations brought different values of $R_0$ with comparable uncertainty, e.g.,
\cite{1972ApJ...174..573H} obtained $R_0=7.0\pm0.6$~kpc,
\cite{1975A&A....41...71O} $R_0=8.7\pm0.6$~kpc,
\cite{1985MmSAI..56...15B} $R_0=6.94\pm0.58$~kpc,
\cite{1986MNRAS.220...69W} $R_0=8.1\pm0.4$~kpc,
\cite{1987MNRAS.226..927F} $R_0=8.0\pm0.65$~kpc,
\cite{1995AJ....110.1674C} $R_0=7.8\pm0.4$~kpc,
\cite{1997MNRAS.284..761F} $R_0=8.1\pm0.4$~kpc,
\cite{2006ApJ...651..197C} $R_0=8.3\pm0.7$~kpc,
\cite{2008A&A...481..441G} $R_0=7.94\pm0.63$~kpc, and recently
\cite{2010AcA....60...55M} $R_0=8.1\pm0.6$~kpc.

In this paper, using the newly released catalog (\citealt{2011AcA....61....1S})
of nearly 17,000 RR Lyr stars detected toward the Galactic bulge during the third
phase of the OGLE project (OGLE-III, \citealt{2003AcA....53..291U,2008AcA....58...69U}),
we show that these variables constitute a uniform population, different to
other populations residing in the central regions of our Galaxy.

The structure of this paper is as follows. Section~\ref{sec:data} describes
the selection of the sample and gives information on the reddening.
Metallicities and magnitudes of the bulge RR Lyr variables are determined
in Section~\ref{sec:metalmags}. In Section~\ref{sec:distance}, we estimate
the distance to the Galactic center. The structure of the RR Lyr population,
based on their mean brightness and density distribution, is studied in
Section~\ref{sec:structure}. Section~\ref{sec:conclusions} states our main
conclusions.


\section{THE DATA}
\label{sec:data}

\subsection{Cleaning the Sample}

\begin{figure}
\centering
\includegraphics[width=8.8cm]{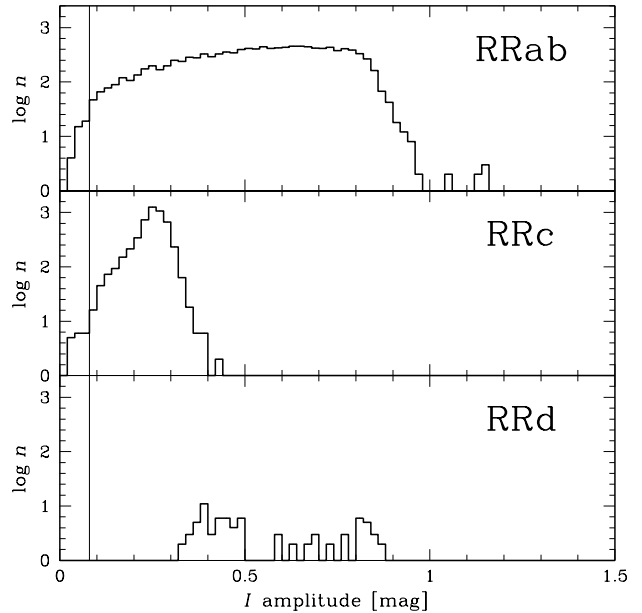}
\caption{Amplitude distributions  for all three
types of RR Lyr variables detected in the OGLE-III bulge area
(\citealt{2011AcA....61....1S}). A bin size of 0.02~mag is used.
Stars with $I$-band amplitudes lower than
0.08~mag (marked with a thin vertical line) are very likely blended objects
and were rejected from the analysis. Note that the histograms show
logarithm of the counts, thus bins with a single star appear blank.}
\label{fig:histamp}
\medskip
\end{figure}

\cite{2011AcA....61....1S} published data on 11,756 RRab, 4989 RRc,
and 91 RRd stars observed toward the Milky Way bulge. The mean 
$I$-band magnitudes of the detected stars lie in the range $12 \leq I \leq 20$.
For the purpose of our work, the sample had to be cleaned from various
contaminants. In the first step we rejected RR Lyr with amplitudes lower than
0.08 mag in $I$, which are very likely blended objects. Figure~\ref{fig:histamp}
presents $I$-band amplitude distributions for each of the three RR Lyr variable
types. Based on the shape of the distributions for RRab and RRc stars we
set the lower limit at an amplitude of 0.08~mag resulting in the rejection of
38 RRab, 17 RRc, and 0 RRd objects. We also removed the object OGLE-BLG-RRLYR-02792,
which likely belongs to a binary system, and a RRab variable with
a suspiciously large amplitude of 2.0~mag. The remaining amplitudes,
of stars left in the sample, were no higher than 1.49~mag.

\begin{figure}
\centering
\includegraphics[width=8.8cm]{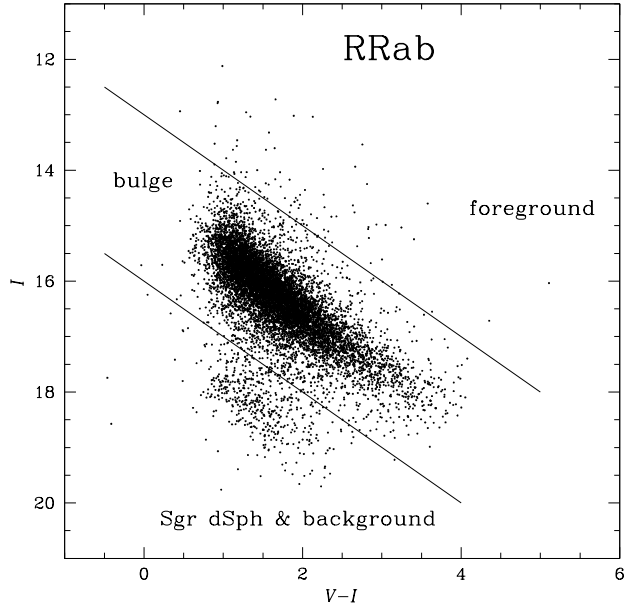}
\caption{Color--magnitude diagram (based on mean values) for 11,124
OGLE-III RRab variables. For further analysis only stars located between
the two inclined lines were used. Most of the background objects belong to the
Sagittarius dwarf spheroidal (Sgr dSph) galaxy. Accuracy of the mean
brightness is estimated as no worse than 0.02 and 0.05~mag in the
$I$ and $V$ bands, respectively.}
\label{fig:iviRRab}
\medskip
\end{figure}

\begin{figure}
\centering
\includegraphics[width=8.8cm]{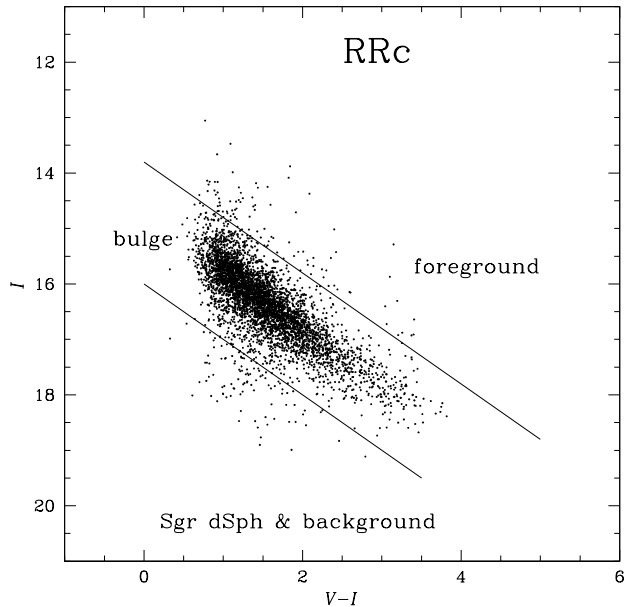}
\caption{Same as in Figure~\ref{fig:iviRRab} but for 4837 RRc variables.}
\label{fig:iviRRc}
\medskip
\end{figure}

\begin{figure}
\centering
\includegraphics[width=8.8cm]{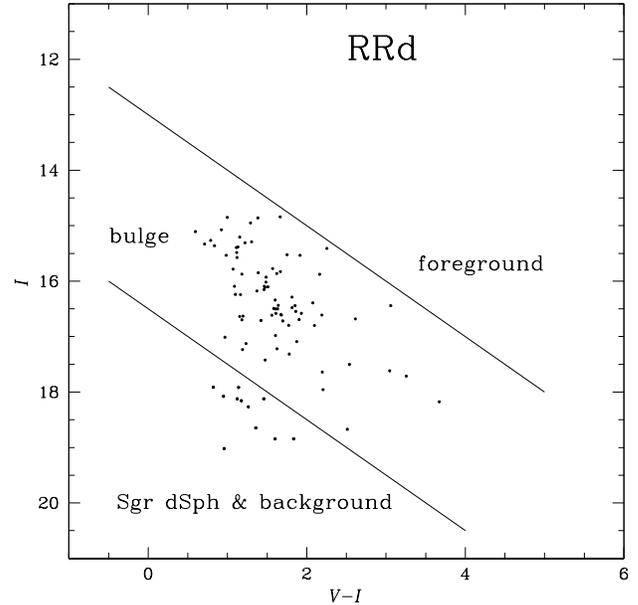}
\caption{Same as in Figures~\ref{fig:iviRRab} and \ref{fig:iviRRc} but for
89 RRd variables.}
\label{fig:iviRRd}
\medskip
\end{figure}

In the next step, we constructed $I$ versus $V-I$ diagrams to separate
bulge RR Lyr stars from foreground and background objects. The diagrams
for RRab, RRc, and RRd variables are shown in
Figures~\ref{fig:iviRRab}--\ref{fig:iviRRd}, respectively. In all these
figures one can see the effect of heavy reddening. Most of the background RR Lyr
stars belong to the Sagittarius dwarf spheroidal (Sgr dSph) galaxy. This is
well seen in the case of numerous and bright RRab stars which form a sequence
in the lower part of Figure~\ref{fig:iviRRab}.

Finally, from the list of bulge RR Lyr stars we rejected 51 RRab and 23 RRc
variables which are members or likely members of eight Galactic globular
clusters (GCs; P. Pietrukowicz et al., in preparation).

\subsection{Reddening}
\label{sec:reddening}

\begin{figure}
\centering
\includegraphics[width=8.8cm]{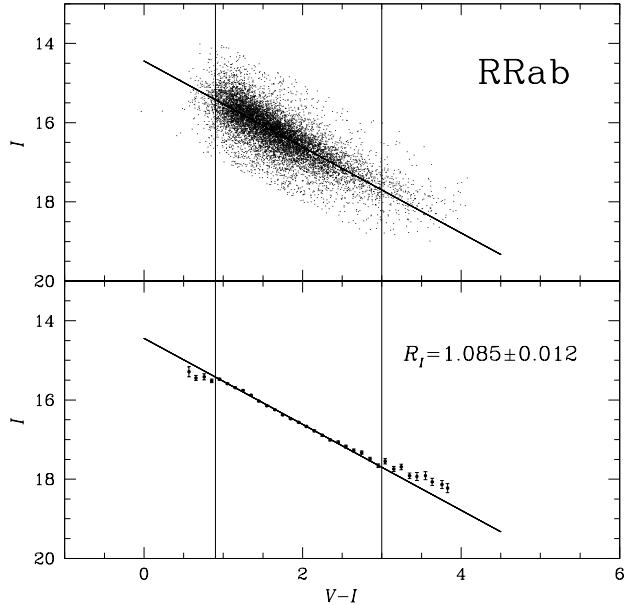}
\caption{Color--magnitude diagram with a linear fit for unbinned (upper
panel) and binned (lower panel) bulge RRab variables. The bin size is
0.1~mag. Thin vertical lines mark section to which we fit the straight
line.}
\label{fig:VVIRRab}
\medskip
\end{figure}

\begin{figure}
\centering
\includegraphics[width=8.8cm]{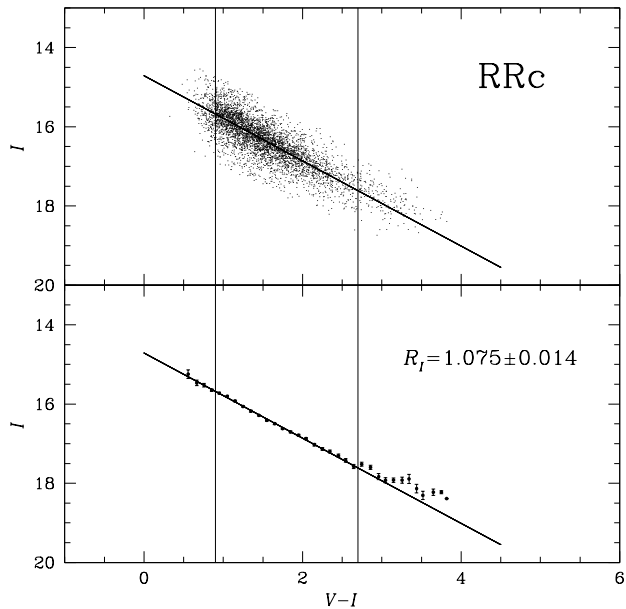}
\caption{Same as in Figure~\ref{fig:VVIRRc} but for RRc type variables.
Note very similar value of $R_I$ for RRab and RRc stars.}
\label{fig:VVIRRc}
\medskip
\end{figure}

The selection returned 10,472 RRab, 4608 RRc, and 78 RRd bulge
variables with measured mean $V$- and $I$-band magnitudes.\footnote{We note
that all magnitudes published in the catalog of \cite{2011AcA....61....1S}
were calibrated to the standard \cite{1992AJ....104..372L} photometric
system using transformation formulae presented in \cite{2002AcA....52..217U}
and an improved relation for stars with $(V-I)>2$~mag in \cite{2011AcA....61...83S}.}
The corrected $I$-band magnitudes are accurate to 0.01-0.02~mag for
$(V-I)<6$~mag. For an additional 627 RRab, 151 RRc, and 2 RRd
stars that are located close to the Galactic plane, we do not
have information on $V$-band brightness due to the heavy reddening
in those directions. We used the information on the $I$-band brightness
and $V-I$ color to estimate the ratio of total to selective extinction
$R_I=A_I/E(V-I)$, and to test whether it is independent of color or not.
In Figures~\ref{fig:VVIRRab} and \ref{fig:VVIRRc}, respectively for RRab and
RRc stars, we show unbinned and color-binned bulge RR Lyr stars in the
$I$ versus $V-I$ diagram. The observed scatter of $\sim0.4$~mag is due to
intrinsic differences in the absolute brightness, differences in extinction
along the line of sight to the RR Lyr star, the effect of the extent of the
population along the line of sight, and the photometric uncertainty.
We mark the limits in $V-I$ between which we fit a linear regression.
Stars with high $V-I$ are very reddened and too faint to be observed.
On the other side, there are very few mildly reddened bulge RR Lyr stars.
We obtained the following relations for RRab and RRc stars, respectively:
\begin{equation}
\label{equ:1}
I_{\rm RRab}=(1.085\pm0.012)(V-I)+(14.443\pm0.021),~~\sigma=0.42,
\end{equation}
\begin{equation}
\label{equ:2}
I_{\rm RRc}=(1.075\pm0.014)(V-I)+(14.712\pm0.022),~~~\sigma=0.36.
\end{equation}
From the fits we find a mean value of $R_I=1.080\pm0.007$.
We also find the difference in mean $I$-band brightness between RRc and
RRab stars to be $\Delta I=0.27\pm0.03$~mag. The obtained value of
the ratio of total to selective extinction is in agreement with $R_I=1.1$
inferred from the OGLE-II data by \cite{2003ApJ...590..284U}.\footnote{A
later determined value of 1.168 by \cite{2010ApJ...721L..28N} from
brightness of OGLE-III red clump stars is slightly different, since they
calibrated photometry for stars redder than $(V-I)=2$~mag using an older and
slightly inaccurate formula.} Our result indicates that interstellar extinction
toward the bulge is indeed anomalous if compared with the standard value
of $R_{I,VI}=1.4$ derived from the all-sky maps of \cite{1998ApJ...500..525S}.

\subsection{Selection of the Subfields}

\begin{figure}
\centering
\includegraphics[width=8.8cm]{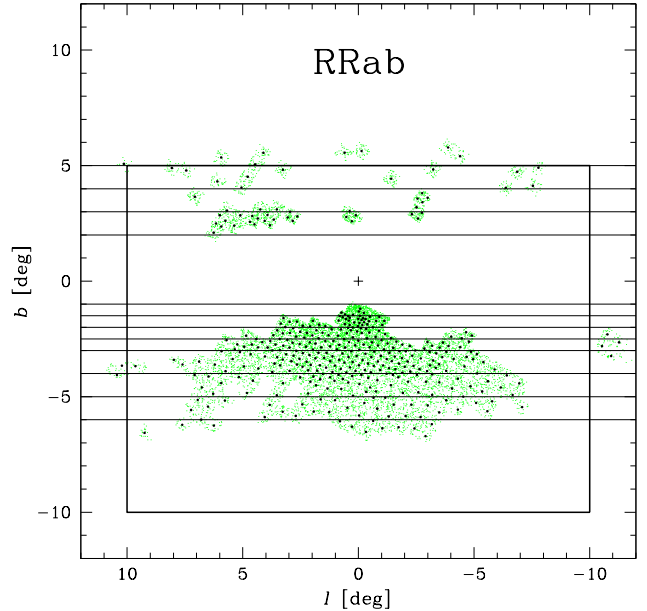}
\caption{Location of 11,099 bulge RRab variables in Galactic coordinates.
Large points denote centers of 396 square subfields. Horizontal lines
limit 10 selected latitudinal stripes used in further analysis.
With the rectangle we mark the bulge area of the VVV
near-IR survey (\citealt{2010NewA...15..433M}).}
\label{fig:mapRRlb}
\medskip
\end{figure}

The field of view of the OGLE-III 8-chip camera was a square $35\farcm6$
on a side. We used this area as a unit field in our analysis. To obtain
good statistics of RR Lyr stars, we divided the unit field
into four and further into sixteen square subfields,
corresponding to two adjacent OGLE-III chips ($17\farcm8$ on a side)
and half of a chip ($8\farcm9$ on a side), respectively. The lowest
number of RR Lyr stars required per subfield we set at 10. If the number
of stars was lower than 10, we counted stars in a larger subfield.
Poorly populated unit fields (with less than 10 stars) were not taken into
account in the analysis of metallicity and brightness distributions
(Section~\ref{sec:metalmags}). In Figure~\ref{fig:mapRRlb}, we show
a map of the bulge RRab stars in Galactic coordinates together with
marked centers of selected subfields. The number of RRab
subfields is 396 with 10--64 stars per subfield. In case of the RRc
stars we selected 200 subfields, each containing 10--53 stars.

\subsection{Completeness of the Sample}

Completeness of the OGLE-III catalog of RR Lyr stars depends on many
factors such as brightness of stars, their amplitude, shape of the
light curve, crowding, and number of observing points. The search for
periodic variability was run on $I$-band light curves with more than
30 points for about $3\times 10^8$ stars with brightness down to
$I\sim21$~mag. The light curves typically consist of 100--3000 measurements.
In order to test and improve completeness of the catalog, \cite{2011AcA....61....1S}
carried out a cross-identification with published catalogs of RR Lyr stars based
on MACHO (\citealt{2008AJ....135..631K}) and OGLE-II data
(\citealt{2006ApJ...651..197C}). Out of 2114 MACHO RRab stars they missed
only 27 objects. Thirteen of them turned out to be located close to bright,
saturated stars that were masked during reductions. The cross-match with
1888 OGLE-II RRab variables resulted in one missing object, a blended star.
Based on the cross-identification, \cite{2011AcA....61....1S} estimated
the completeness of the bulge RRab variables at a level of about 99\%.

\begin{figure}
\centering
\includegraphics[width=8.8cm]{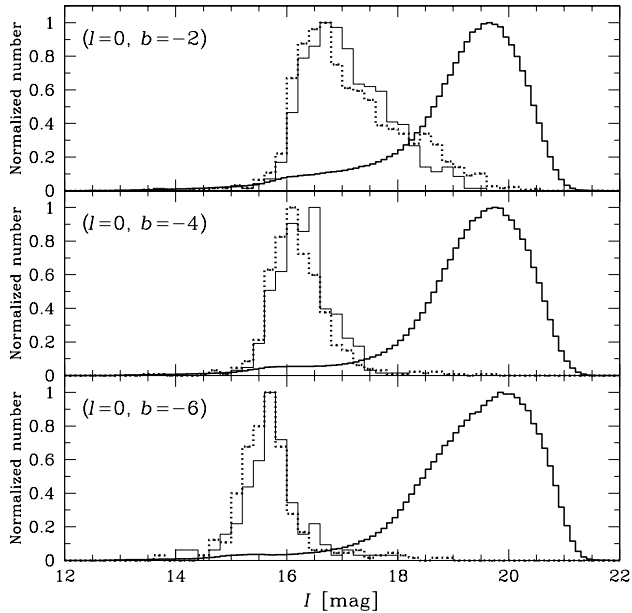}
\caption{Luminosity functions of bulge RRab (dotted line), RRc (thin solid line),
and all stars (thick solid line) detected in the OGLE-III $I$-band data
within $1\fdg0$ around $(l,b)=(0,-2)$ (upper panel), $(l,b)=(0,-4)$
(middle panel), and $(l,b)=(0,-6)$ (lower panel). Note that the OGLE
photometry is complete down to $I\sim19.5$~mag, which is deeper than
the mean brightness of almost all detected RR Lyr stars.}
\label{fig:lumfunallRRI}
\medskip
\end{figure}

In Figure~\ref{fig:lumfunallRRI}, we compare the luminosity histograms for all detected
OGLE-III RRab and RRc variables with the histograms for all stars detected stars
toward three different Galactic latitude intervals. The OGLE photometry
is complete down to $I\sim19.5$~mag and there are very few RR Lyr stars
with mean magnitudes fainter than this brightness. This shows that the
limiting magnitude weakly affects the total number of detected RR Lyr stars. 
The completeness, at least for bright RRab stars with characteristic
high-amplitude saw-tooth-shaped light curves, is indeed very high.
RRc stars, on the other hand, are on average fainter than RRab variables
and have low-amplitude sinusoidal light curves that are very similar to
the light curves of ellipsoidal binaries. Hence, pulsating stars of this
type are more difficult to classify.

\section{METALLICITY AND BRIGHTNESS OF THE BULGE RR LYRAE STARS}
\label{sec:metalmags}

\subsection{Metallicity}

According to the idea of \cite{1995A&A...293L..57K}, the iron abundance
of RRab stars can be derived from their light curve parameters.
The method was further developed by \cite{1996A&A...312..111J} who
presented a basic relation between the metallicity $\feh$, pulsational
period $P$, and Fourier phase combination $\phi_{31}=\phi_3-3\phi_1$
from $V$-band light curves. We determine metallicities for 10,259 reliable
bulge RRab variables using a relation suitable for $I$-band light curves
found by \cite{2005AcA....55...59S} for sine decomposition:
\begin{equation}
\label{equ:3}
{\rm \feh_{J95}} = - 3.142 - 4.902 P + 0.824 \phi_{31},~~\sigma_{\rm sys}=0.18.
\end{equation}
The derived metallicities are on the \cite{1995AcA....45..653J} scale (J95),
which is based on high dispersion spectroscopic measurements. We used the
uncertainty of $\phi_{31}$ to calculate the intrinsic metallicity uncertainty
$\sigma_{\rm int}$ of each star. They range from 0.002
to 0.30~dex with a median value of 0.012~dex. This uncertainty quadratically
added to $\sigma_{\rm sys}$ given by \cite{2005AcA....55...59S} results in a
median total uncertainty of $\sigma_{\rm [Fe/H]}=0.18$~dex.

\begin{figure}
\centering
\includegraphics[width=8.8cm]{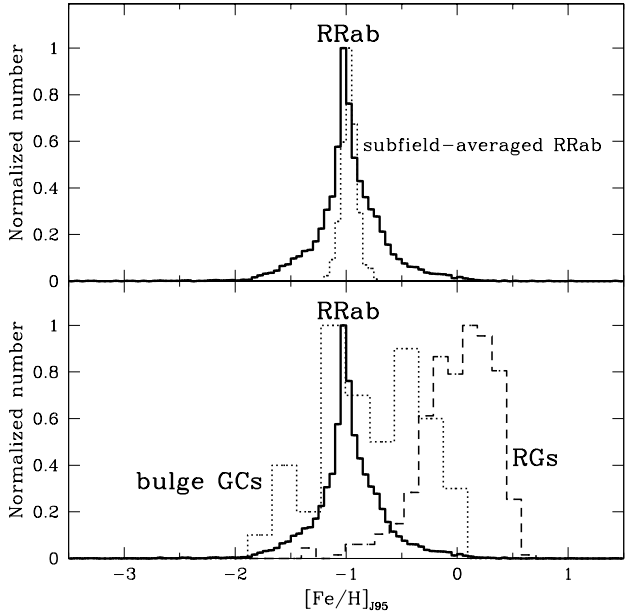}
\caption{Normalized metallicity distribution (on the \cite{1995AcA....45..653J}
scale) for 10,259 bulge RRab stars (solid thick line) in comparison
with the distribution of average RR Lyr metallicity for the 396
subfields that were analyzed (upper panel) and in comparison with
distributions for 47 bulge GCs located within 3.0~kpc
from the Galactic center, and 403 bulge red giants (RGs) from
the catalog of \cite{2008A&A...486..177Z} (lower panel).
The distribution of RR Lyr stars is sharply peaked at $\feh=-1.02$~dex
and significantly differs from the populations of the bulge RGs and GCs.}
\label{fig:histmetalRRab}
\medskip
\end{figure}

In Figure~\ref{fig:histmetalRRab} we show the metallicity distribution
for the bulge RRab stars, plotted as a histogram with a bin size of
0.05~dex. In the upper panel of this figure we plot, for comparison, a
distribution of average RR Lyr metallicity for the 396 subfields which
were analyzed. Both distributions are sharply peaked, which indicates
that the bulge RR Lyr population was likely formed on a short timescale, as it is
in the case of the Galactic halo and thick disk (\citealt{1996ASPC...92..307G}).
About 98.7\% of the bulge RRab stars have metallicities between $-2.0$
and 0.0~dex, with a peak at $\feh_{J95}=-1.02$ and a dispersion
of 0.25~dex. These numbers are in very good agreement with results
obtained by \cite{2008AJ....136.2441K} from $V$-band light curves of
2435 MACHO RRab variables. They found an average metallicity of
$\langle\feh\rangle=-1.25$ and a dispersion of 0.30~dex on the
\cite{1984ApJS...55...45Z} metallicity scale which, according to the
work of \cite{2000AJ....119..851P}, is less metal-rich by about 0.24~dex
than the scale of \cite{1995AcA....45..653J}.

\begin{figure}
\centering
\includegraphics[width=8.8cm]{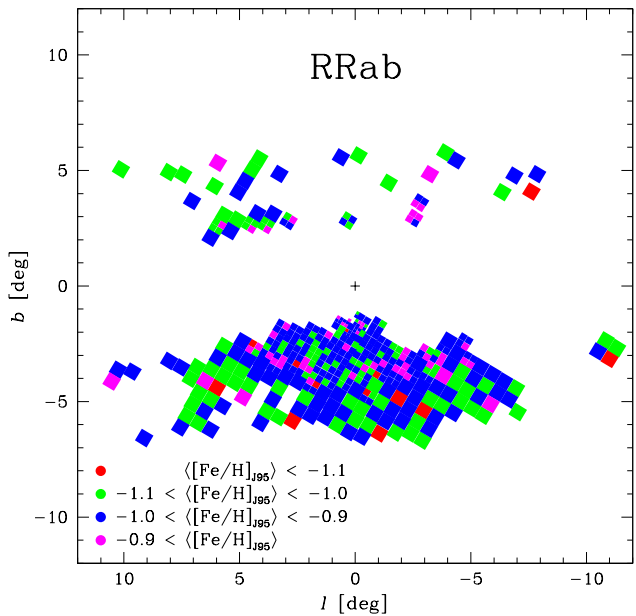}
\caption{Metallicity map for RRab variables averaged in 396 analyzed
subfields. The metallicity appears to be uniformly distributed over the
observed part of the Galactic bulge.}
\label{fig:metalmaplb}
\medskip
\end{figure}

In the lower panel of Figure~\ref{fig:histmetalRRab}, we compare the bulge
RRab metallicity-distribution function derived in this work with that of
the 47 Galactic globular clusters (GCs) located within 3.0~kpc from
the Galactic center\footnote{According to \cite{2001AJ....122.2587C} and
the updated Catalogue of Variable Stars in Galactic Globular Clusters (2011) by
C.~Clement (http://www.astro.utoronto.ca/$\sim$cclement/read.html) the 47 bulge
GCs contain 325 known RR Lyr stars considered to be members of the clusters.
Around two-thirds of those variables (exactly 217 stars) belong to the globular
cluster M62. There are no known RR Lyr stars in 31 of the selected bulge GCs.}
(based on 2010 version of \citealt{1996AJ....112.1487H} catalog) and that
of a sample of 403 bulge red giants (RGs) published by
\cite{2008A&A...486..177Z}. The metallicities of the GCs and RGs were
converted from the \cite{2009A&A...508..695C} scale (C09) to the one of
\cite{1995AcA....45..653J} via the \cite{1984ApJS...55...45Z}
metallicity scale (ZW84) using a relation from \cite{2000AJ....119..851P}:
\begin{equation}
\label{equ:4}
{\rm \feh_{ZW84}} = 1.028 {\rm \feh_{J95}} - 0.242,
\end{equation}
and a second relation from \cite{2009A&A...508..695C}:
\begin{equation}
\label{equ:5}
{\rm \feh_{C09}} = 1.105 {\rm \feh_{ZW84}} + 0.160.
\end{equation}
The metallicity distributions of the bulge RGs, RR Lyr stars, and GCs
differ significantly from each other. The very peaked and symmetric distribution
for the RRab stars indicates that the population of the bulge RR Lyr
variables is very metal uniform. In Figure~\ref{fig:metalmaplb} we show
a color-coded map of metallicity averaged in each of the 396 subfields.
The metallicity of RR Lyr stars seems to be fairly uniformly distributed
over the whole investigated area.

\subsection{Dereddened Mean Brightness}
\label{sec:deredmags}

We calculate dereddened mean magnitudes $I_0$ for each RR Lyr star as:
\begin{equation}
\label{equ:6}
I_0 = I - A_I,
\end{equation}
where the extinction is given by
\begin{equation}
\label{equ:7}
A_I = R_I E(V-I),
\end{equation}
with reddening given by
\begin{equation}
\label{equ:8}
E(V-I) = (V-I) - (V-I)_0.
\end{equation}
We compute the intrinsic color $(V-I)_0$ for RRab stars as
\begin{equation}
\label{equ:9}
(V-I)_0 = M_V - M_I,
\end{equation}
where the absolute brightnesses $M_V$ and $M_I$ come from theoretical
calibrations in \cite{2004ApJS..154..633C}
\begin{equation}
\label{equ:10}
M_V = 2.288 + 0.882~\log Z + 0.108~(\log Z)^2,
\end{equation}
\begin{equation}
\label{equ:11}
M_I = 0.471 - 1.132~\log P + 0.205~\log Z,
\end{equation}
with the following conversion between $Z$ and $\feh$:
\begin{equation}
\label{equ:12}
{\rm log} Z = {\rm [Fe/H]} - 1.765.
\end{equation}
The conversion is based on a solar metallicity of $Z_{\odot}=0.01716$,
as required to match the metal-to-hydrogen ratio $Z/X$ of
\cite{1993oee..conf...15G}. Intrinsic colors of RRc stars are between
0.32 and 0.42~mag (based on models in \citealt{1999A&A...351..103F}).
Since the metallicites of this type of RR Lyr variables are not determined,
we use the mean value of $(V-I)_0=0.37$~mag for each RRc star.

In the above calculation, we adopt an uncertainty of the mean $I$-band
brightness at a level of $\sigma_I=0.02$~mag (\citealt{2008AcA....58...69U}).
The OGLE-III $V$-band light curves are less well sampled and the adopted
accuracy of the mean $V$-band brightness is $\sigma_V=0.05$~mag.
The errors propagate to a range of $\sigma_{I0}=0.07$--0.20~mag for the
sample of 10,259 RRab stars with a median value of 0.09~mag.


\section{DISTANCE TO THE GALACTIC CENTER}
\label{sec:distance}

\begin{figure}
\centering
\includegraphics[width=8.8cm]{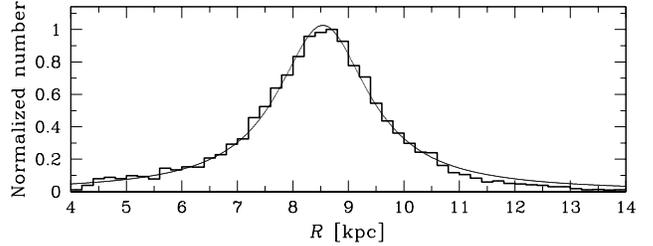}
\caption{Distance distribution for 10,259 bulge RRab stars.
The bin size is 0.2~kpc. The distribution is very well represented
by a Lorentzian with a maximum value at $R_0=8.54$~kpc.}
\label{fig:histdist}
\medskip
\end{figure}

We estimate the position of the Galactic center by measuring the mean
distance to the bulge RR Lyr variables. The distances $R$
to individual bulge RRab star are estimated using the relation:
\begin{equation}
\label{equ:13}
\log R = 1 + 0.2 (I_0 - M_I).
\end{equation}
To obtain the real distance to the center of the population we have
to apply the following two geometric corrections. First, we have to project
all the individual distances onto the Galactic plane by taking the cosine
of the Galactic latitude, yielding $R{\rm cos}b$. Second, we have to take
into account the ``cone-effect'' --- our subfields subtend solid angles
on the sky with more volume further away --- by scaling their distance
distribution by $R^{-2}$. A corrected and normalized distance distribution
for the 10,259 RRab stars is presented in Figure~\ref{fig:histdist}.
The symmetry of this distribution further confirms the high completeness
of the RR Lyr catalog of \cite{2011AcA....61....1S}. The slightly
flat peak in the distribution between 8.2 and 8.8~kpc is consistent with
the presence of an elongated inner structure (see Section~\ref{sec:structure}).
The distance distribution is very well represented by a Lorentzian with
a maximum value at $8.540\pm0.013$~kpc. A resulting distance to the
Milky Way center of $R_0=8.54\pm0.42$~kpc is obtained by quadratically
adding the median uncertainty in the distance to the individual stars of
0.418~kpc. This result is in good agreement with the weighted average
of $8.15\pm0.14\pm0.35$~kpc determined from eleven different measuring
methods by \cite{2010RvMP...82.3121G}.

\section{STRUCTURE OF THE BULGE RR LYRAE POPULATION}
\label{sec:structure}

\subsection{Brightness Distribution}

\begin{figure}
\centering
\includegraphics[width=8.8cm]{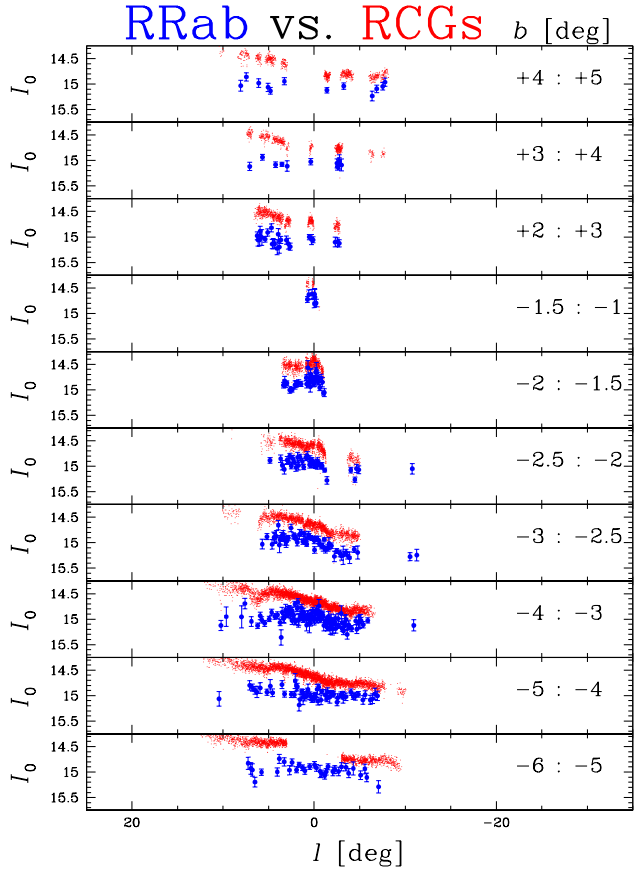}
\caption{Dereddened $I$-band magnitude distribution for bulge RRab stars
(blue points), in comparison with red clump giants (RCGs, red points)
in 10 latitudinal stripes (marked in Figure~\ref{fig:mapRRlb}). Not shown
are the individual errors on the magnitudes of RCGs, which are between
0.10 and 0.40~mag, with a median value of 0.20~mag. The RR Lyr
stars only follow the distribution of RCGs toward the inner part of the
Galactic bulge ($-3\degr<l<+3\degr$, $-4\degr<b<-2\degr$).}
\label{fig:I0sliceRRab}
\medskip
\end{figure}

\begin{figure}
\centering
\includegraphics[width=8.8cm]{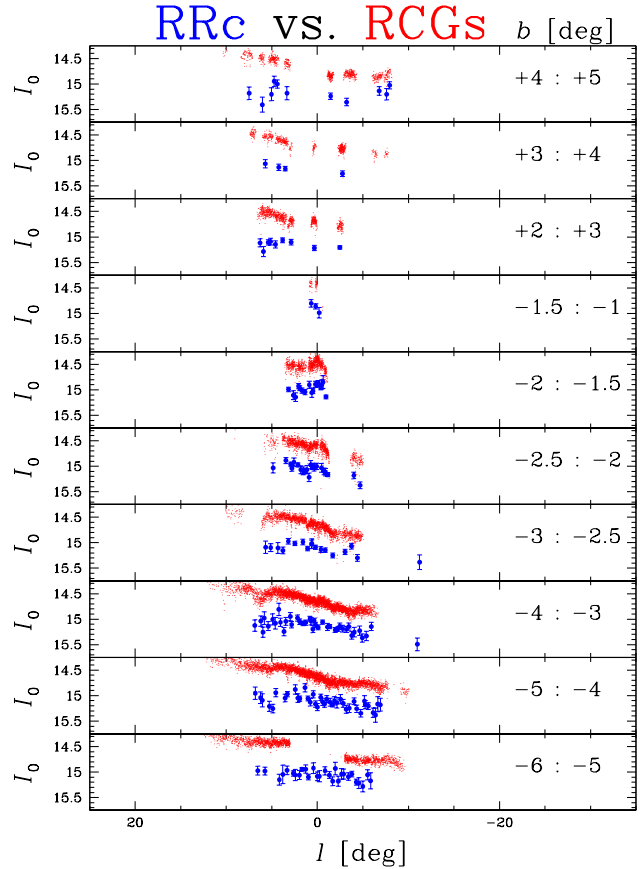}
\caption{Same as in Figure~\ref{fig:I0sliceRRab} but for the bulge RRc stars.}
\label{fig:I0sliceRRc}
\medskip
\end{figure}

\begin{figure}
\centering
\includegraphics[width=8.8cm]{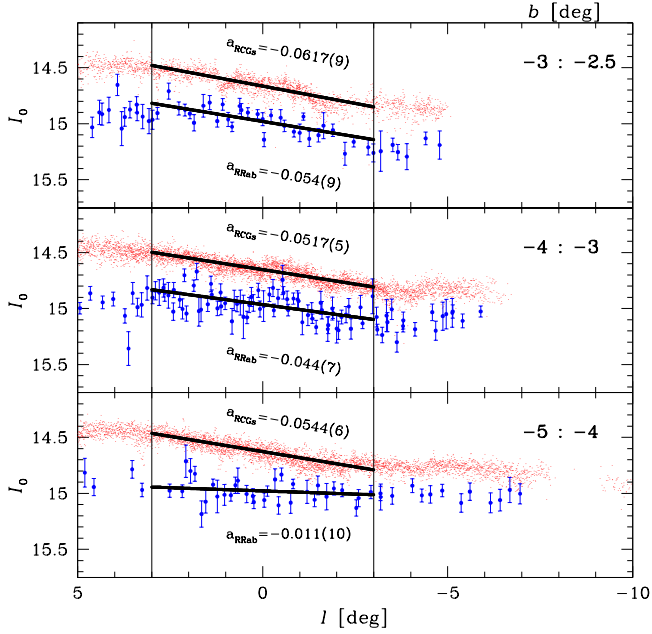}
\caption{Comparison of the inclinations $a$ for RRab stars (blue points)
and RCGs (red points) in relation $I_0$ vs. $l$ for the Galactic longitudes
$-3\degr<l<+3\degr$ in three well-observed latitudinal sections. Uncertainties
of the inclinations are given in parenthesis. In the inner regions
($|b|<4\degr$) RR Lyr stars more closely follow the trend of the RCGs.}
\label{fig:I0fitsRRab}
\medskip
\end{figure}

In this section, we study the structure of the bulge RR Lyr population
using their mean dereddened magnitudes at different Galactic latitudes.
We selected 10 latitudinal stripes, they are $0\fdg5$ or $1\fdg0$ wide.
Their location is illustrated in Figure~\ref{fig:mapRRlb}.
Figure~\ref{fig:I0sliceRRab} shows $I_0$ magnitudes averaged in subfields
for RRab stars in comparison to the mean brightnesses of RCGs from
D. M. Nataf et al. (2012, in preparation). A typical uncertainty in the
dereddened brightness of the RCGs is 0.20~mag. Results for RRc stars
are shown in Figure~\ref{fig:I0sliceRRc}. Bulge RR Lyr stars follow
the barred RCGs distribution in the inner part of the bulge at
$|l|<3\degr$ and $|b|<4\degr$. This is well seen in the better-covered
southern part (see Figure~\ref{fig:I0fitsRRab}). Farther off the
Galactic plane at latitudes $|b|\gtrsim4\degr$ the distribution
of RR Lyr variables is clearly flat, indicating that their population
becomes more spherical with increasing distance from the center.

\begin{figure}
\centering
\includegraphics[width=8.8cm]{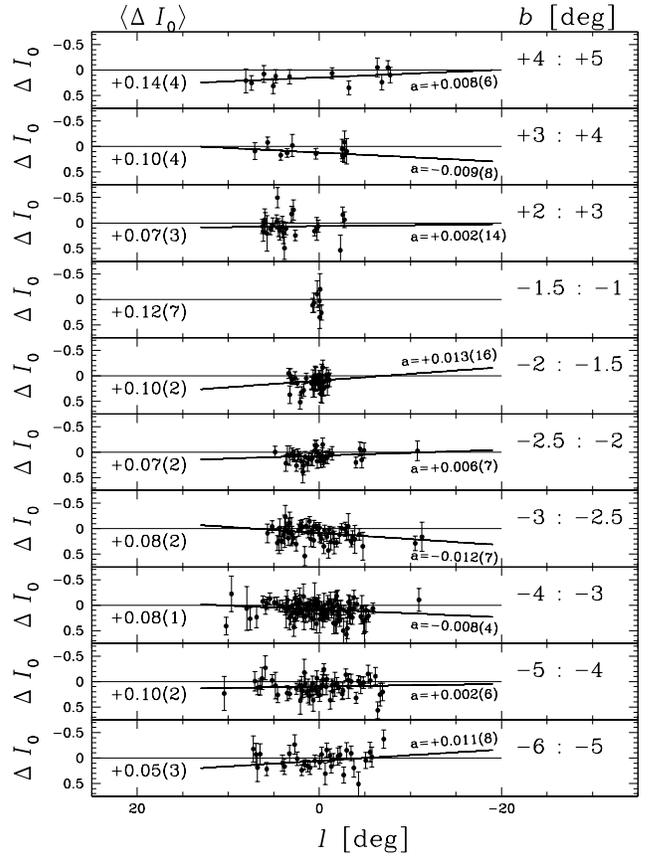}
\caption{Differences in dereddened $I$-band magnitudes between metal-rich
($\feh>-1.0$) and metal-poor ($\feh<-1.0$) RRab stars in 10 latitudinal
stripes. Mean differences $\langle \Delta I_0 \rangle$ for all stripes
are around a similar value of $\sim+0.1$~mag and there are no significant
trends in $\Delta I_0$ with Galactic longitude, as is expected for a
metal-uniform population.}
\label{fig:diffI0sliceRRab}
\bigskip
\end{figure}

In Figure~\ref{fig:diffI0sliceRRab}, we investigate if there are any
differences in dereddened magnitudes between metal-rich ($\feh_{\rm J95} > -1.0$)
and metal-poor ($\feh_{\rm J95} < -1.0$) RRab stars. For each latitudinal
stripe we fit a weighted linear regression to the data.
All derived mean values are positive, as expected from Equation~(\ref{equ:11}),
that metal-poor stars are on average brighter than metal-rich ones.
All values are around $+0.1$~mag, indicating no significant differences
in metallicity distribution along the line of sight of the bulge RR Lyr
population for different Galactic latitudes. Moreover, there seems to be
no significant trends in $\Delta I_0$ with Galactic longitude. These
results corroborate the metal-uniform nature of the bulge RR Lyr population.

\subsection{The Inner Part}

\begin{figure}
\centering
\includegraphics[width=8.8cm]{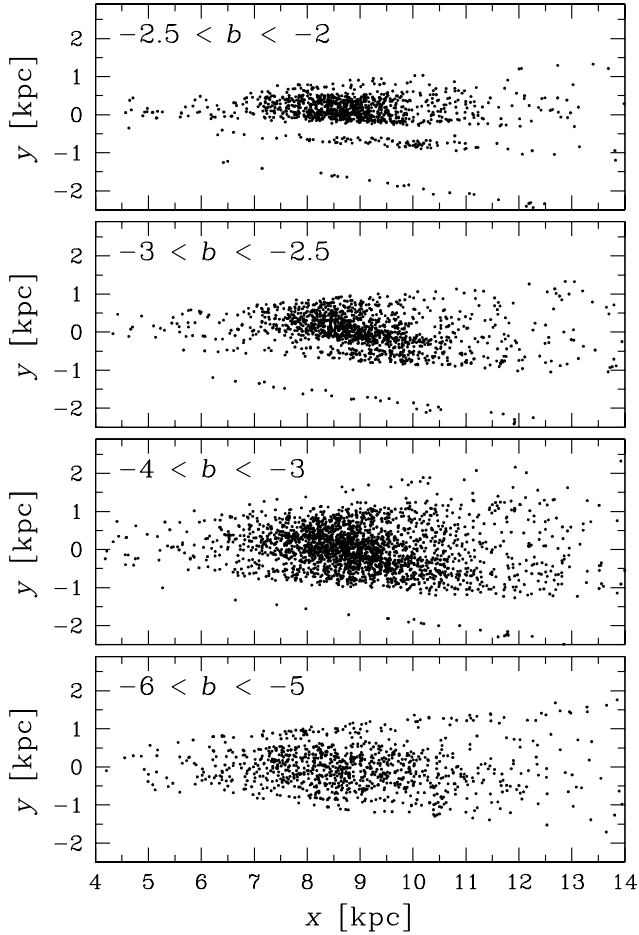}
\caption{RRab variables from four different latitudinal stripes
projected onto the Galactic plane, where the Sun is located at $(x,y)=(0,0)$.
For stars located closer to the plane the elongated structure is evident.
It disappears farther off the plane. Note that the coverage of the bulge
area at some longitudes is incomplete (see Figure~\ref{fig:mapRRlb})}
\label{fig:RRabpro}
\medskip
\end{figure}

\begin{figure}
\centering
\includegraphics[width=8.8cm]{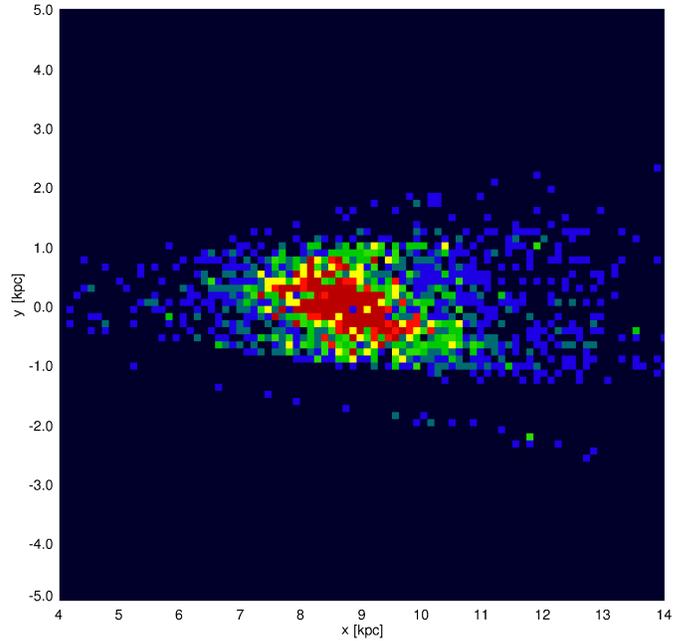}
\caption{Density map of RRab stars with latitudes $-4\degr<b<-3\degr$
projected onto the Galactic plane. Each bin is 0.12~kpc on a side.
The inner structure is elongated and inclined by $\sim30\degr$
with respect to the line of sight.}
\label{fig:mappro}
\medskip
\end{figure}

The OGLE-III $V$- and $I$-band data cover Galactic latitudes higher than
$b\sim+2\fdg1$ and lower than $b\sim-1\fdg2$. Using RRab
variables with derived individual distances, we can make an attempt to study
the shape and orientation of the observed part of the inner bulge.
Figure~\ref{fig:RRabpro} shows four projections of the variables onto
the Galactic plane. Two of those projections, $-3\degr<b<-4\degr$
and $-6\degr<b<-5\degr$, include complete data for stars with Galactic
longitudes $-4\degr<l<+4\degr$. Figure~\ref{fig:mappro} illustrates the
projection for $-3\degr<b<-4\degr$ as a color--density map. The presence of
the elongated structure for the RR Lyr located closer to the Galactic plane
is evident. Farther off the plane the structure gets round, as is expected
from the flat mean magnitude distribution presented in
Figures~\ref{fig:I0sliceRRab}--\ref{fig:I0fitsRRab}.
Based on the density map in Figure~\ref{fig:mappro}, we find the inclination of
the RR Lyr structure major axis to be about $30\degr$ with respect to the line
of sight between the sun and the Galactic center. This angle is very
similar to the inclination of $24\degr$--$27\degr$
derived for the main RCGs bar by \cite{2007MNRAS.378.1064}. However,
we stress that the accurate estimation of the inclination of the RR Lyr
spatial distribution is hampered due to small number statistics and incomplete
coverage of the central part of the bulge. Near-IR data from the on going
VISTA Variables in the Via Lactea (VVV) survey (\citealt{2010NewA...15..433M})
should bring the final answer to the question on the structure and shape
of the inner bulge RR Lyr population.

\subsection{Search for Split in Brightness of the RR Lyr Stars}

\begin{figure}
\centering
\includegraphics[width=8.8cm]{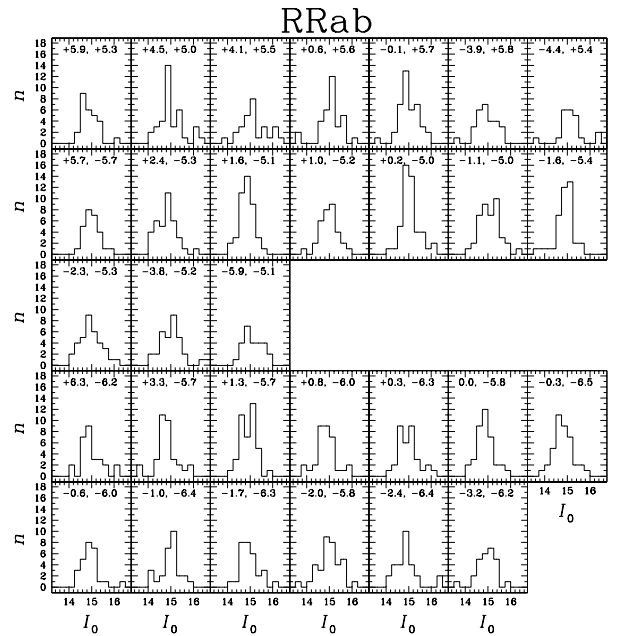}
\caption{Number histograms of RRab stars in selected subfields around
Galactic latitudes $+5\degr$ (upper row), $-5\degr$
(upper middle and middle rows), and $-6\degr$ (two bottom rows).
Galactic coordinates for the centers of the subfields are given
in each box. The size of the bins is 0.25~mag, i.e., almost three times
wider than a typical photometric error ($\sigma_{I0}\sim0.09$~mag). 
Small number statistics prevent us from concluding whether a split
such as that observed for RCG exists for RR Lyr.}
\label{fig:splitI0}
\medskip
\end{figure}

Recently, \cite{2010ApJ...721L..28N} used OGLE-III photometry to find that the
Galactic bulge red clump is split into two components toward the fields
at ($-3\fdg5<l<1\degr$, $b<-5\degr)$ and $(l,b)=(0,+5\fdg2)$. The split is
also reported by \cite{2011A&A...534A...3G} who analyzed the first data from
the VVV survey (\citealt{2012A&A...537A.107S}). The observed difference
in $I$-band brightness between the components is approximately 0.5~mag.
We selected OGLE fields located at $|b|>5\degr$ for which the number
of detected bulge RRab stars was at least 20. Figure~\ref{fig:splitI0}
presents luminosity histograms for the investigated fields.
The adopted size of the bins, 0.25~mag, is wider than the typical
brightness uncertainty of $\sigma_{I0}=0.08$~mag. The number of bulge
RR Lyr variables is insufficient to conclude if such a split is present
in this type of stars or not. Some histograms display double peaks, but
this can be explained by small number statistics.

\subsection{Density Distribution}

\begin{figure}
\centering
\includegraphics[width=8.8cm]{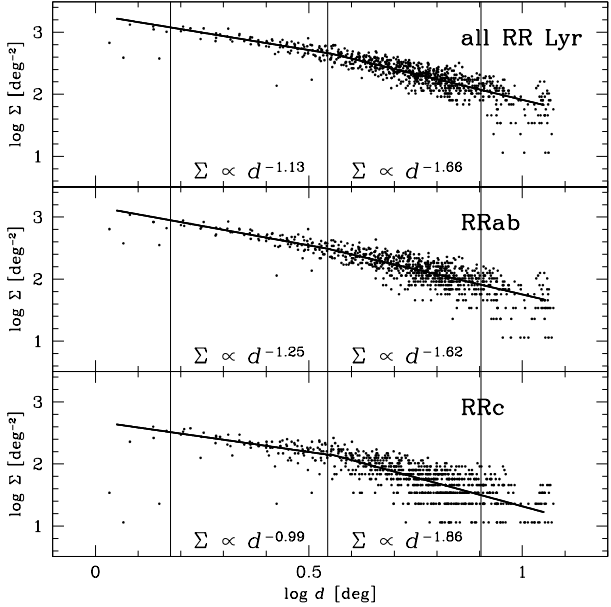}
\caption{Mean surface density distribution as a function of angular
distance from the Milky Way center for all bulge RR Lyr (upper panel),
RRab (middle panel), and RRc variables exclusively (lower panel).
Three vertical lines at $\log d=0.176$, 0.544, and 0.903, corresponding,
respectively, to linear distance $r=0.2$, 0.5, and 1.2~kpc, mark two
sections of the data to which lines were fit. The outliers at
$\log d\approx0.25$, $0.42$, and $0.51$ were not included in the
fitting for the density relations.}
\label{fig:densityRR}
\medskip
\end{figure}

We analyzed the mean spatial distribution of the bulge RR Lyr as
a function of distance from the Milky Way center by counting the stars
in square subfields that are $17\farcm8$ on a side. Separate results for
the whole sample (consisting of all RR Lyr pulsators), RRab, and RRc
subsamples are presented in three panels of Figure~\ref{fig:densityRR}.
The data cover angular distances between $d\approx1\fdg1$
and $d\approx11\fdg8$ from the center, however, the innermost
subfields are incomplete due to high reddening, while the
outermost ones suffer from small number statistics. We found a break
in both distributions at $d\approx3\fdg5$. For all detected RR Lyr
variables in the inner part between $d=1\fdg5$ and $d=3\fdg5$
we obtained the following fit:
\begin{equation}
\label{equ:15}
\log \Sigma_{\rm allRR,inn} = (-1.13\pm0.08) \log d + (3.28\pm0.04),
\end{equation}
where $d$ is in degrees and $\Sigma$ in counts deg$^{-2}$.
The outer part for $3\fdg5<d<8\fdg0$ is represented by a
steeper line with
\begin{equation}
\label{equ:14}
\log \Sigma_{\rm allRR,out} = (-1.66\pm0.06) \log d + (3.57\pm0.04).
\end{equation}
In the case of RRc stars we found, respectively,
\begin{equation}
\label{equ:17}
\log \Sigma_{\rm RRc,inn} = (-0.99\pm0.13) \log d + (2.69\pm0.06)
\end{equation}
and:
\begin{equation}
\label{equ:16}
\log \Sigma_{\rm RRc,out} = (-1.86\pm0.10) \log d + (3.18\pm0.07).
\end{equation}
Finally, for RRab stars we have
\begin{equation}
\label{equ:19}
\log \Sigma_{\rm RRab,inn} = (-1.25\pm0.09) \log d + (3.17\pm0.04)
\end{equation}
and
\begin{equation}
\label{equ:18}
\log \Sigma_{\rm RRab,out} = (-1.62\pm0.07) \log d + (3.37\pm0.05).
\end{equation}
This would correspond to a spherical distribution with
$$
D_{\rm RRab,inn} \propto r^{-2.25\pm0.09}~~~{\rm for}~~0.2<r<0.5~{\rm kpc}
$$
and
$$
D_{\rm RRab,out} \propto r^{-2.62\pm0.07}~~~{\rm for}~~0.5<r<1.2~{\rm kpc},
$$
assuming the distance to the Galactic center of $R_0=8.54$~kpc given
in Section~\ref{sec:distance}.

The distribution of RR Lyr stars in the central part ($r<0.2$~kpc) remains
unknown. We may speculate that it probably gets more flat toward the
center. The answer may come from $K_s$-band observations currently
collected by the VVV survey (\citealt{2010NewA...15..433M}).
Using the OGLE-III data, we are able to
assess the number of RR Lyr stars which should be detected within the VVV
bulge area (marked in Figure~\ref{fig:mapRRlb}). If we extrapolate
the inner distribution toward the center, we find that VVV should detect
$(60\pm7)\times10^3$ RRab stars. With a flat distribution in the central part
we obtain $(49\pm7)\times10^3$ stars. For RR Lyr variables of all types
the numbers are the following: $(82\pm9)\times10^3$ and $(71\pm9)\times10^3$.
In summary, if we assume that the OGLE-III catalog of RRab variables
is 99\% complete, we can estimate the VVV survey should detect
$(4$--$7)\times10^4$ RR Lyr stars of this type.

\begin{figure}
\centering
\includegraphics[width=8.8cm]{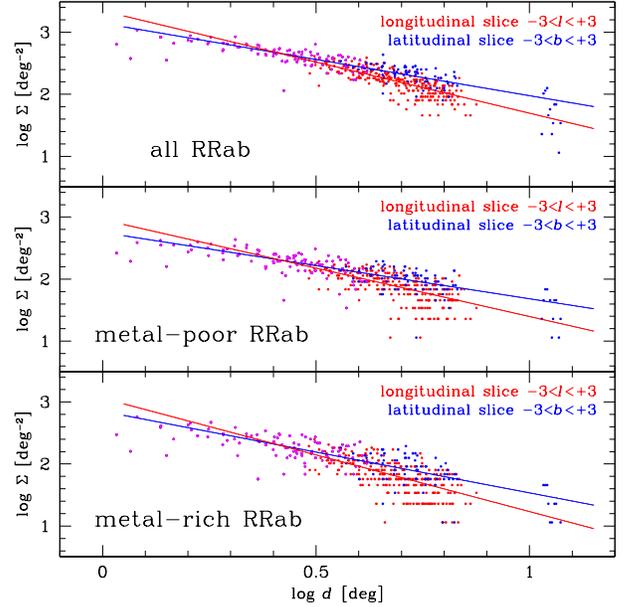}
\caption{Mean surface density distributions for RRab stars of all
metallicities (upper panel), metal-poor ($\feh_{\rm J95}<-1.0$, middle panel),
and metal-rich ($\feh_{\rm J95}>-1.0$, lower panel) located within $6\degr$ wide
longitudinal (in red) and latitudinal slices (in blue). All points
with $\log d<0.477$ (in magenta) are common. There is no major difference
between the three distributions, which further confirms the metal-uniform
nature of the bulge RR Lyr population. The steeper slope for the longitudinal
slice indicates that the population is flattened along Galactic plane.}
\label{fig:densityRRab}
\medskip
\end{figure}

In Figure~\ref{fig:densityRRab} we draw density distributions of bulge RRab
variables in longitudinal and latitudinal slices, each with width of $6\degr$,
to verify if there are any significant differences between metal-poor
($\feh_{\rm J95}<-1.0$) and metal-rich ($\feh_{\rm J95}>-1.0$) stars.
No differences are found in the linear fits, leading us to the earlier
conclusion that the RR Lyr bulge population is metal uniform.
The observed slope of the latitudinal distributions is always shallower
than the longitudinal one, confirming earlier suggestions of
\cite{1998IAUS..184..123M} that the whole population is slightly flattened
along Galactic longitude. Based on the longitudinal and latitudinal
fits obtained we find a mean density of ${\rm log}\Sigma=2.2$ at distances
$d_{\rm long}=4\fdg9$ (${\rm log}d_{\rm long}=0.69$) and $d_{\rm lat}=6\fdg4$
(${\rm log}d_{\rm lat}=0.81$), respectively. From this we can derive that
the RR Lyr surface distribution is flattened in the observed outer part,
with $b/a=d_{\rm long}/d_{\rm lat}\sim0.75$. Complete data from the
near-IR VVV survey will allow to study the distribution in the very
central part of the bulge and to determine final parameters for the
bulge RR Lyr population.

\section{CONCLUSIONS}
\label{sec:conclusions}

We have analyzed the data on 16,836 RR Lyr stars observed toward the Galactic
bulge during the third phase of the OGLE project (years 2001--2009).
After eliminating low-amplitude stars, objects likely foreground
and background sources, and GC members the sample includes 10,472 RRab,
4608 RRc, and 78 RRd bulge variables with derived mean $V$- and $I$-band
brightness. For an additional 627 RRab, 151 RRc, and 2 RRd stars, all located
close to the Galactic plane, we possess mean magnitudes in the $I$ filter alone.

Using color information for this type of standard candles, we find that
the ratio of total to selective extinction $R_I=A_I/E(V-I)$ is independent
of color and equals $1.080\pm0.007$. This result confirms
the anomalous nature of the extinction toward the bulge.

Our analysis clearly demonstrates that the bulge RR Lyr stars constitute
a very uniform population, different to the majority of the
stars in the central regions of the Milky Way. The photometrically
derived metallicity distribution is sharply peaked at
$\feh_{\rm J95}=-1.02\pm0.18$ with a dispersion of 0.25~dex on
\cite{1995AcA....45..653J} metallicity scale. Based on the distribution
of dereddened brightness we show that the RR Lyr population is elongated
in the inner part, where it tends to follow the barred distribution of the bulge
RCGs. The elongated structure can also be noticed as a flat peak in the
distance distribution. A possible explanation of this structure along
the bar is that less populated old RR Lyr stars may feel gravitational
forces from young stars in the Milky Way bulge. Farther off the
Galactic plane (for $|b|\gtrsim4\degr$) the shape of RR Lyr population
becomes spherical. We note that, due to small number statistics, we can
not conclude the presence of a split in the population of bulge
RR Lyr stars, as is observed in RCGs.

The distance to the Milky Way center inferred from the OGLE-III bulge
RRab variables is $R_0=8.54\pm0.42$~kpc. This result is in
agreement with $R_0=8.1\pm0.6$~kpc obtained from the OGLE-II data
(\citealt{2010AcA....60...55M}) and with a value
of $R_0=8.15\pm0.49$~kpc combined by \cite{2010RvMP...82.3121G}
from 11 different methods. It seems that the RR Lyr method suffers
from uncertainties in the metallicity determination, and hence absolute
brightness of the stars.

In our work, we have also studied the density distribution of the bulge
RR Lyr stars as a function of distance from the Galactic center.
We found a break in the distribution at a distance of $\sim0.5$~kpc
indicating flattening toward the center. Taking into
account completeness of the OGLE-III data, we estimate the number
of RRab stars which should be detected within the bulge area
of the near-IR VVV survey as $(4$--$7)\times10^4$ stars.
The uncertainty partially results from unknown density distribution
within $0.2$~kpc from the center.

On going and future wide-field photometric surveys, such as OGLE-IV, VVV,
Pan-STARRS (\citealt{2002SPIE.4836..154K}), and Large Synoptic Survey
Telescope (\citealt{2008ApJ...684..287I}) should bring
definitive answers to questions pertaining to the structure and evolution
of all Galactic RR Lyr populations. For example, one of unsolved
issues is whether the bulge and halo RR Lyr variables form the same
population. \cite{1998IAUS..184..123M} suggested that the bulge
RR Lyr stars could represent the inner extension of the Galactic
halo. The bulge pulsators do not show the \cite{1939Obs....62..104O}
dichotomy observed in the halo variables. From the period distribution
and period--amplitude diagram presented in \cite{2011AcA....61....1S}
it seems that probably all bulge RR Lyr stars belong to the Oosterhoff
type I (OoI). Our work shows that the bulge population is very metal
uniform and was likely formed on a short time scale. The Galactic
halo was probably formed by at least two distinct accretion processes
(\citealt{2008ApJ...678..865M}). About 25\% of the halo RR Lyr
variables observed at distances 3--30~kpc appear to be Oosterhoff type
II (OoII) stars with a significantly different radial density
profile to the remaining 75\% of OoI objects.
The OoI halo population has a power-law exponent of $-2.26\pm0.1$
while the OoII component has a steeper slope with $-2.88\pm0.04$
(\citealt{2008ApJ...678..865M}). The OoI stars are on average
slightly more metal-rich ($\langle\feh_{\rm J95}\rangle\sim-1.7$~dex)
than the OoII stars ($\langle\feh_{\rm J95}\rangle\sim-2.0$~dex;
\citealt{2010ApJ...708..717S}). Comparison of the above numbers
for the halo RR Lyr variables with those for the bulge objects obtained
in this work does not allow us to answer the question on the common
bulge/halo population. More data, in particular on RR Lyr stars
residing in the central and outer regions of our Galaxy, are needed.


\acknowledgments

We thank M\'arcio Catelan, Wojciech Dziembowski, Andrew Gould, Dante Minniti,
and the anonymous referee for useful comments that helped to improve the paper.
The OGLE project is supported by the European Research Council under
the European Community's Seventh Framework Programme (FP7/2007-2013),
ERC grant agreement no. 246678 to A.U. P.P. is also supported
by the Grant No. IP2010 031570 financed by the Polish Ministry of
Sciences and Higher Education under Iuventus Plus program.


\end{document}